\documentclass[12pt]{amsart}

\usepackage{graphicx} 

\usepackage{verbatim} 

\usepackage[usenames,dvipsnames]{color}

\usepackage{hyperref} 

\usepackage{amsmath,amsthm,amsfonts}

\usepackage{morefloats}

\usepackage{fullpage}

\usepackage {ulem}

\newtheorem{thm}{Theorem}[section]

\theoremstyle{definition}

\newtheorem{definition}[thm]{Definition}

\theoremstyle{remark}

\newtheorem{rem}[thm]{Remark}

\numberwithin{equation}{section}

\newcommand{\CVA}{\text{CVA}}
\newcommand{\BCVA}{\text{BCVA}}
\newcommand{\UCVA}{\text{UCVA}}
\newcommand{\PCVA}{\text{PCVA}}
\newcommand{\FTDCVA}{\text{FTDCVA}}

\newcommand{\DVA}{\text{DVA}}
\newcommand{\UDVA}{\text{UDVA}}
\newcommand{\PDVA}{\text{PDVA}}
\newcommand{\FTDDVA}{\text{FTDDVA}}

\newcommand\commt[1]{}

\begin{document}

\pagestyle {plain}

\title{Restructuring Counterparty Credit Risk}%

\author{Claudio Albanese}

\email{claudio.albanese@global-valuation.com}%
\address{Global Valuation Limited, London, EC2M 4YF, UK}

\author{Damiano Brigo}%

\email{damiano.brigo@kcl.ac.uk}
\address{Dept. of Mathematics, King's College, London, Strand, London, WC2R 2LS }

\author{Frank Oertel}%

\email{frank.oertel@bafin.de}
\address{Dept. of Cross-Sectoral Risk Modelling (Q RM), Federal Banking Supervisory Authority (BaFin), D-53117 Bonn}

\thanks{We thank Carla di Mauro, Fernanda D'Ippoliti, Giacomo Pietronero and Gary Wong for comments and discussions. Views expressed in this work are the authors' views and are not necessarily shared by the Federal Financial Supervisory Authority (BaFin). This paper was submitted to appear in the Deutsche Bundesbank Discussion Paper Series. We are thankful to an anonymous referee for the careful work and useful comments.}

{}%




\date{April 13, 2012}%




\begin{abstract}

We introduce an innovative theoretical framework to model derivative transactions between defaultable entities based on the principle of arbitrage freedom.  Our framework extends the traditional formulations based on Credit and Debit Valuation Adjustments (\CVA\; and \DVA).

Depending on how the default contingency is accounted for, we list a total of ten different structuring styles. These include bipartite structures between a bank and a counterparty, tri-partite structures with one margin lender  in addition, quadri-partite structures with two margin lenders and, most importantly, configurations where all derivative transactions are cleared through a Central Counterparty (CCP). 

We compare the various structuring styles under a number of criteria including consistency from an accounting standpoint, counterparty risk hedgeability, numerical complexity, transaction portability upon default, induced behaviour and macro-economic impact of the implied wealth allocation.

\end{abstract}

\maketitle


\tableofcontents

\section{Introduction}

Whenever a defaultable entity enters into a financial transaction, it sustains a cost of carry to compensate counterparties for its potential inability to meet future contractual obligations. The compensation mechanism for counterparty credit risk is captured by a protection contract contingent to the default arrival time and to the exposure at default. In the particular case of debt transactions, the cost of carry of the contingent default protection liability is the cost of funds above the riskless rate. In the case of derivative contracts, the cost of carry of default protection is either captured by a Credit Valuation Adjustment (\CVA), or by the cost of posting collateral, or a combination thereof.

A number of different structuring styles for counterparty credit risk have so far been proposed and several market standards have emerged. Nevertheless, renewed efforts to arrive at a stable market standard are still under way, particularly as counterparty credit risk management is an element of key importance to the financial reforms enacted on the wake of the 2007 banking crisis.

Traditionally, \CVA\; risk has been retained on banks' balance sheets along with the risk implied by the CVA mark-to-market volatility. The CVA risk is transferred internally within banks from business desks to CVA desks by transacting a Contingent Credit Default Swap (CCDS). A CCDS offers (to the protection buyer) protection against default of a reference entity, the nominal being given by the valuation of a reference portfolio at time of default. The valuation of the CVA for derivative books is described in \cite{ABGP2011} while individual CCDS contracts are analyzed in \cite{BrigoPallavicini2008}. 

An open market for CCDS contracts never established itself. The Financial Times reported, back in 2008 in \cite{Cookson}, that
\begin{quote}
Rudimentary and idiosyncratic versions of these so-called contingent credit default swaps (CCDS) have existed for five years, but they have been rarely traded due to high costs, low liquidity and limited scope.
\end{quote}
As a consequence, counterparty credit risk has traditionally been managed with bipartite structures, whereby two parties agree to exchange mutual default protection without the intervention of third parties. However, as we discuss in this paper, the move 
towards CCP clearing in OTC markets opens new venues for counterparty credit risk transfer through multi-party arrangements. 

\subsection{From unilateral CVA and CCDS to bilateral CVA}

Prior to 2007, counterparty credit risk was accounted for in terms of the so called unilateral CVA (or UCVA), a valuation methodology which stems from the modeling premise that the party carrying out the valuation is default free. See for example \cite{BrigoMasetti} for the general framework under netting, and \cite{BrigoPallavicini2007}, \cite{BrigoBakkar}, \cite{BrigoChourdakis}, \cite{BrigoMoriniTarenghi} for applications of this framework to various asset classes.

The UCVA is faulty as the valuation is asymmetric between the two parties and breaks the basic accounting principle according to which an asset for one party represents a liability for the counterparty. Only monetary authorities have the authority to evade this accounting principle, which we refer to as {\it money conservation}. 
Accounting standards such as FAS 156 and 157 remedied the inconsistency by prescribing that banks have to mark a unilateral Debt Valuation Adjustment (UDVA) which is equal in amount to the unilateral CVA entry realized by the counterparty but sits on the opposite side of the double entry ledger (cf. \eqref{eq_v_t_c}). The CVA and DVA thus give contributions of opposite signs to the portfolio Net Asset Value (NAV). 

This reflects ``money conservation'' and applies to all forms of CVA including unilateral CVA and the other variants described below such as first-to-default CVA and portable CVA. In the unilateral case, the combination
\begin{equation}
\BCVA : = \UCVA - \UDVA 
\end{equation}
defines the so called bilateral CVA, see for instance \cite{Picoult}. The omission of the UDVA term is justified only when one of the two parties in the trade can be considered as being default-free, an assumption that was easily granted to a number of financial institutions prior to 2007.

While restoring money conservation, accounting for the DVA introduced another valuation inconsistency as close-out conditions ignored the UDVA and the loss of UDVA upon counterparty default was not properly priced and accounted for. To lessen the impact of the inconsistency, in 2009 the ISDA modified the wording of the close-out rule in the Credit Support Annex (CSA) portion of master swap agreements, see \cite{ISDA2009}. Under the new rule, the unilateral DVA is recoverable through the liquidation process on the same footing and with the same seniority and recovery rate of the mark-to-market valuation of the underlying transaction. Since the UDVA is not recoverable entirely but only in part, this modification roughly halves the impact of the mispricing and accounting inconsistency, without however eliminating it entirely. 

In this paper, we consider a number of alternative contractual specifications which instead are fully consistent from an accounting viewpoint and correctly price the possible loss of a fraction of DVA.

\subsection{First-to-default CVA and First-to-default DVA}

The first consistent structures introduced historically are bilateral versions of the CVA and DVA which include a first-to-default contingency clause and are denoted with FTDCVA and FTDDVA. The first-to-default clause appears in \cite{DuffieHuang}, \cite{BieleckiRutkowski2002} and was made explicit in \cite{BrigoCapponi2010} in the case of an underlying CDS. See also \cite{BrigoMorini2011} and
\cite{BrigoMorini2010Flux}. It was considered in the case of interest rate portfolios in \cite{BrigoPallaviciniPapatheodorou} and also appears in \cite{Gregory2009}. In this context, the paper \cite{BrigoCapponiPallaviciniPapatheodorou} extends the bilateral theory to collateralization and re-hypothecation and \cite{BrigoCapponiPallavicini} shows cases of extreme contagion where even continuous collateralization does not eliminate counterparty risk.

The first-to-default CVA is consistent from an accounting standpoint. However, this definition does not come without shortcomings.  
One problem is that, if the credit of the computing party is low, the FTDCVA can be substantially lower than the UCVA. At default, the first-to-default CVA even vanishes. These material discrepancies between the first-to-default CVA and the unilateral CVA contemplated in the Basel III Accord have undesirable side effects, such as for instance giving a competitive advantage to those financial institutions which are the slowest at endorsing the banking reform.   In \cite{WattRisk}, we read
\begin{quotation}
"I would say the top 10 banks in the market are all aware of this, are discussing it between themselves and are making the appropriate pricing changes",
says Christophe Coutte, global co-head of flow fixed income and currencies at Soci\'et\'e G\'en\'erale Corporate and Investment Banking (SG CIB) in London. "However, there are many smaller, regional banks that are not fully charging for the extra CVA capital in derivatives transactions. As we've prudently stepped up our prices, the tier-two and tier-three firms have filled the gap we've left. They weren't competing before because they couldn't offer enough liquidity or the tightest bid-offer spreads, but now they can." 
\end{quotation}

A second problem is that the \textit{first-to-default} CVA is unhedgeable since, to hedge it, a bank would have to short its own credit, an impossible trade. This aggravates the already serious problem of unhedgeability of the unilateral DVA term as it appears in the definition of the BCVA. Finally, if the first-to-default CVA is used for pricing while the unilateral CVA is used to determine capital requirements, the material mismatch between the two makes it difficult to optimize risk management strategies. Furthermore, if a bank charges the FTDCVA to clients but then has to provision a higher amount given by the UCVA, the difference needs to be provisioned either by separate fundraising or from Tier 1 capital. 

\subsection{Basel III}
The Basel III Accord prescribes that banks should compute unilateral CVA by assuming independence of exposure and default. The advanced framework allows banks to implement the effect of wrong way risk (i.\,e., the risk that occurs when the bank's exposure to its counterparty is adversely correlated with the credit quality of that counterparty) in the calculation of their exposures by using own models, while under the standardized approach the Basel III Accord accounts for the effect by means of a one-size-fits-all multiplier. Examples in \cite{BrigoPallavicini2007}, \cite{BrigoBakkar}, \cite{BrigoChourdakis}, \cite{BrigoMoriniTarenghi} indicate that the actual multiplier (the so called ``alpha'' multiplier) is quite sensitive to model calibration and market conditions and that the advanced framework is more risk sensitive.

Interestingly, the Basel III Accord chooses to ignore the UDVA in the calculation for capital adequacy requirements. Although consideration of the UDVA needs to be included for accounting consistency, no such principle exists as far as capital requirements are concerned.

\subsection{Portable CVA}

Due to the variety of possible different definitions of CVA (unilateral CVA, bilateral CVA, first-to-default CVA) combined with boundary conditions such as risk free or replacement closeouts and the option to use either exact or approximate treatment of wrong way risk, there appears to be material discrepancies in CVA valuation across financial institutions. This was pointed out recently in the article \cite{WattRisk}. The most significant discrepancy is between UCVA and FTDCVA. Both are used in the industry for valuation, while only the former is endorsed for regulatory purposes in Basel III.

In this paper, we introduce consistent structures called portable CVA and DVA (or PCVA and, respectively, PDVA). These structures have not been discussed previously in the literature and are introduced here because we find them interesting from the point of view of facilitating novations in case of a default of clearing members clearing their OTC trades through a CCP. 

From a quantitative standpoint, the difference between UCVA and PCVA is minimal. The portable CVA assessed by an entity converges to the unilateral CVA whenever the credit-worthiness of the entity tends to infinity (i.\,e., the entity is non-defaultable) and in the opposite direction when the entity defaults. The name "portable" reflects the fact that, upon counterparty default, the PDVA reduces exactly to the unilateral DVA. Portable CVA and Portable DVA are bilateral structures in that they are contingent on the credit process of both parties. They are also more conservative than the unilateral structures in the sense that the PCVA and PDVA amounts are slightly larger or at most equal to the UCVA and UDVA, i.\,e., $PCVA\ge UCVA$. 

As we discuss in Section \ref{sec_portable_cva}, an approximate upper bound for the difference between the portable CVA and the unilateral CVA is given by the product between UDVA and the probability of default. Unlike the situation in the case of the FTDCVA, this difference is nearly immaterial in most circumstances.

\subsection{CVA VaR and periodic resets}

The management of CVA volatility risk is crucial to mitigate the CVA VaR capital charge. A proposal is to trade long dated swaps with an embedded obligation for both parties to restructure the trade to equilibrium periodically while exchanging a payment to settle the mark-to-market difference at the reset date. However, this sort of modified swap is not economically equivalent to the original long-dated swaps. 

A not very practical but conceptually interesting alternative is to structure CVA and DVA so that they are reset periodically to equilibrium leading to a floating rate variant of CVA and DVA. As a simple example, let's consider the case of a bipartite transaction between the default-free bank B and the defaultable counterparty C. Instead of charging CVA upfront at the start date for the entire maturity of the transaction, the bank may require a CVA payment at the start date for protection over the first 6 months period. After 6 months, the bank would require a CVA payment for protection for a further six months period and so on, up to the final maturity of the trade. 

The benefit of having periodic resets would be that this would mitigate the volatility of the CVA by forcing it periodically to equilibrium. However, the difficulty in implementing such a structure is that fixing the CVA premium in future semi-annual protection windows would be subject to model risk and could be the cause of legally unresolvable controversy between the two parties. Although there is no regulatory framework surrounding this case yet, there is also no precedent (to our knowledge) of bilateral contracts with contingent fixings which are influenced by one of the two parties' discretionary decisions regarding modeling choices. 

A more legally robust implementation of the idea of floating rate CVA payment is to let third parties intervene competitively in the process. The floating rate structures in the next subsection includes the case of full collateralization and gives examples of such solutions where fixings are controlled by offer and demand. 

\begin{figure}[t]
\begin{center}
\includegraphics[width = 11cm]{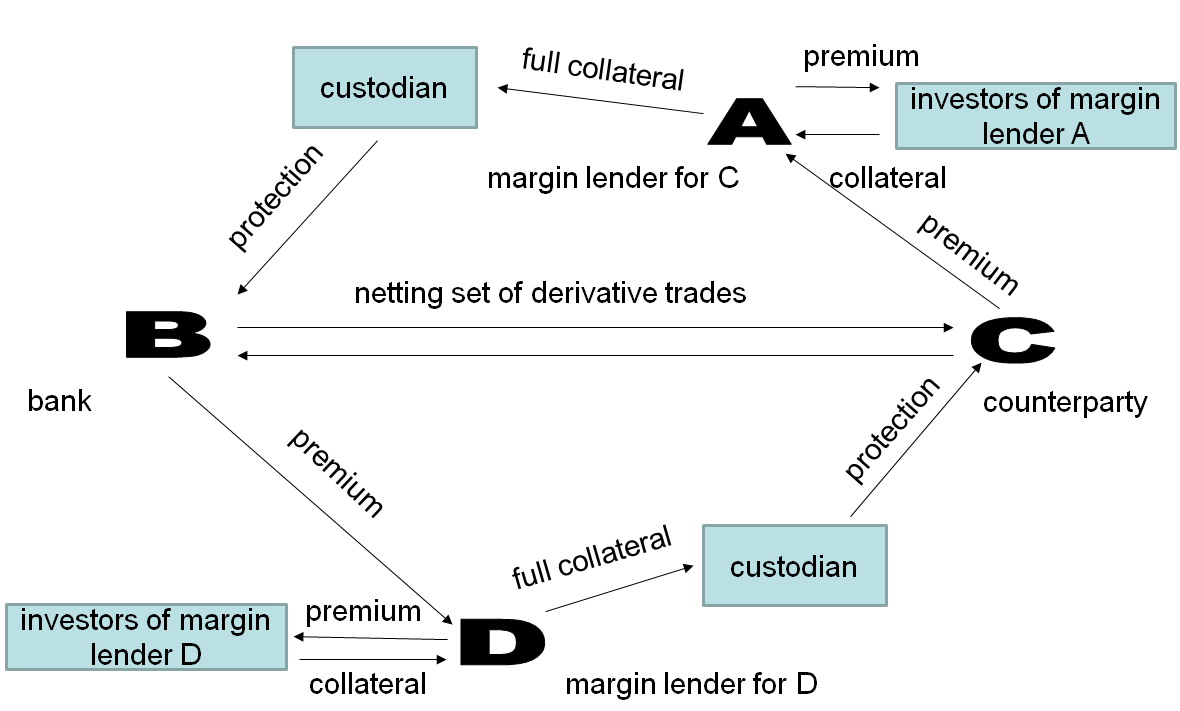}
\end{center}
\caption{General quadri-partite scheme including as special cases bipartite arrangements (just $B$ and $C$), tri-partite arrangements (such as $A$, $B$ and $C$) and quadri-partite arrangements with all four. The margin lenders $A$ and $D$ post collateral.}\label{fig_abcd}
\end{figure}

\subsection{Full collateralization and margin lending}

New classes of structuring styles emerge whenever one introduces the notion of full collateralization and margin lending, see also \cite{APW2011}, \cite{AP2011} and \cite{ADP2011}. Full collateralization is known to be a very effective remedy to reduce counterparty risk. Because of the existence of gap risk and to counteract the volatility of mark-to-market valuations, one actually requires haircuts and overcollateralization schemes to reduce the risk to almost nil, see again \cite{BrigoCapponiPallavicini}. 

The incentive towards full collateralization is built into the regulatory framework itself. The recent article \cite{WattRisk} reports that uncollateralised trades are expected to consume up to four times more capital under Basel III.  The same article reports of the case of Lufthansa, saying 
\begin{quote}
The airline's Cologne-based head of finance, Roland Kern, expects its earnings to become more volatile - not because of unpredictable passenger numbers, interest rates or jet fuel prices, but because it does not post collateral in its derivatives transactions.
\end{quote}

Traditionally, the CVA is typically charged by the structuring bank $B$ either on an upfront basis or by the sake of being built into the structure as a fixed coupon stream. Margin lending instead is predicated on the notion of floating rate CVA payments with periodic resets and is designed in such a way to transfer the credit spread volatility risk and the mark-to-market volatility risk from the bank to the counterparties.
We may explain this more in detail by looking at Figure \ref{fig_abcd}. 

Collateral is most effectively allocated by margin lenders not by transfering the ownership of qualifying collateral assets, but in the form of hypothecs on qualifying collateral. A hypothec is a contract written on underlying assets that transfers neither the ownership nor the possession of the assets, but simply a lien contingent to specific default events occurring. 

The counterparty C is concerned about the amount of collateral she may have to post periodically in order to trade derivatives with bank B. To avoid posting collateral, C enters into a margin lending transaction. C pays periodically (say semi-annually) a floating rate CVA to the margin lender A (`premium' arrow connecting C to A). In turn, the margin lender A distributes the premium to investors according to a seniority hierarchy (premium arrow connecting A to the investors). In exchange for this premium, the investors provide the margin lender A with transferable hypothecs to qualifying collateral assets over a six-months period (`collateral' arrow connecting the investors to A). In turn, A passes the hypothecs to a custodian (`collateral' arrow connecting A to the custodian) to meet collateral calls from the counterparty B. If C defaults within the semi-annual period, B has a claim to the hypothec posted at the custodian and A apportions the loss to its investors in increasing order of seniority, e.g. first to the equity investor, then to the mezzanine, then to the senior, then to the super-senior, etc. In this fashion, A provides protection to B on its derivative exposure with respect to C (`protection' arrow connecting the custodian to B) and the default risk is transferred to the investors of the margin lender on whose collateral the hypothecs had been written.

At the end of the six months period, the protection contract needs to be renewed, thus forcing a reset to an updated premium that keeps the structure at equilibrium. At the end of the period, the margin lender has no residual obligation and may decide whether to bid for a renewal or abstain from it. In turn, C may opt to renew the protection contract with the same margin lender or another or even to syndicate the required coverage across a number of margin lenders. Thanks to the periodic resets, the counterparty C is bearing the CVA volatility risk, whereas B is not exposed to it. This is in stark contrast to the situation which could happen in the case of traditional upfront or fixed rate CVA charges.   

The various structuring styles that can be devised for margin lending are predicated on the separation between default risk on one side, market risk and spread volatility risk on the other. As a consequence, these structures are in general multi-partite. A tri-partite structure involves only one margin lender and only one party that posts full collateral. In a quadri-partite structure, collateral posting obligations are symmetric and there are two lenders, one for the buy-side party and one for the sell-side party. Structuring can also be penta-partite in case there is a Central Counterparty (CCP) that stands as a universal counterparty to both sides. From a contractual viewpoint, multi-party structures still consist of a series of two-party contracts, hence there are no complexities associated to multi-party legal agreements.  

In this paper, we first consider three examples of structures with margin lending: a quadri-partite one with daily resets, a quadri-partite one with periodic resets over longer time intervals and a tri-partite one also with periodic resets. All these structures are consistent from an accounting standpoint and the principle of arbitrage freedom. 

\begin{figure}[t]
\begin{center}
\includegraphics[width = 18cm]{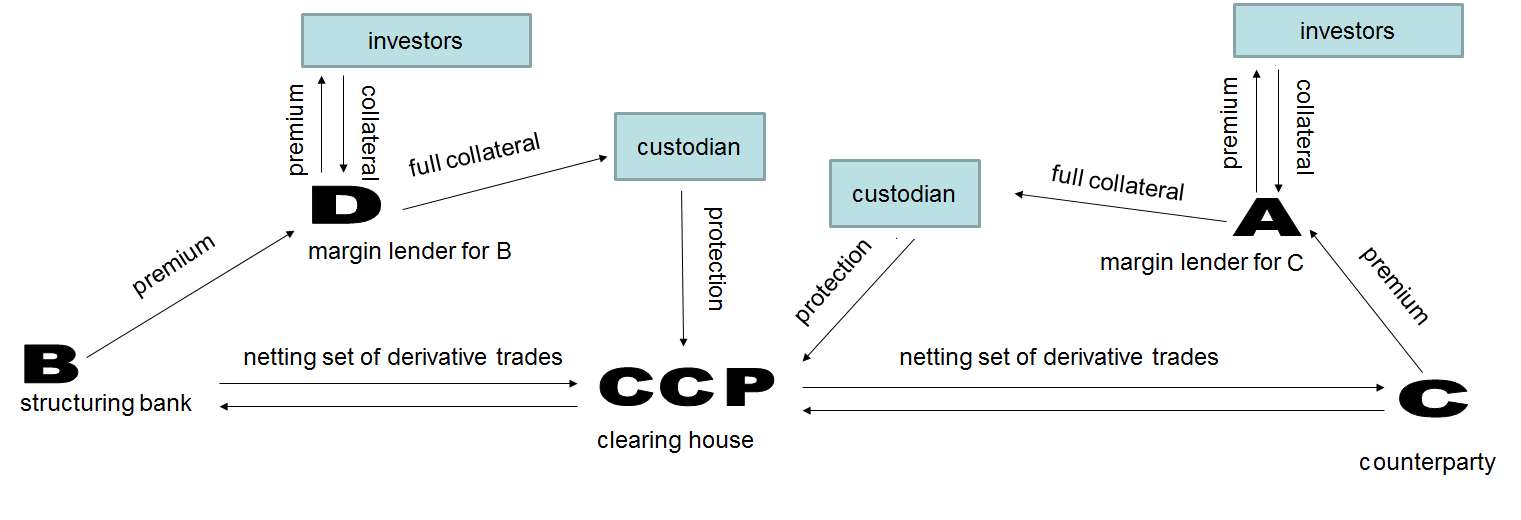}
\end{center}
\caption{Multi-party scheme involving a Central Counterparty.} \label{fig_ccp}
\end{figure}

Being based on floating CVA payments that reset periodically and are proportional to the counterparty's conditional credit spreads and open exposure, quadri-partite structures remedy the unhedgeability issue that plague all the various forms of bipartite CVA by effectively ensuring that CVA volatility risk is absorbed by the party that is responsible for generating it. Default risk instead is passed on to the investors who finance the margin lenders. We suggest this structuring style gives rise to a more resilient market infrastructure than the traditional one based on bilateral long-term CVA and DVA structures which are left to the bank to hedge. If in addition CCPs are present, the effectiveness of margin lending benefits of the greater degree of netting which reduce the carry cost of derivative replication strategies, see Figure \ref{fig_ccp}. 

\subsection{Macro-economic impact}

The traditional CVA and DVA bipartite structures are based on the mutual exchange of long-term default protection contracts.  This effectively embeds credit exposures in each and every derivative transaction. Given the size of global derivative markets, the embedded credit optionality effectively transfers wealth on a global scale with a substantial macro-economic impact. 

All participants in derivative markets, by transacting interest rate swaps or FX options or any other derivative, credit-linked or not, have acquired automatically credit protection on themselves and sold credit protection on their counterparties. This implies that whenever an entity's credit worsens, it receives a subsidy from its counterparties in the form of a DVA credit protection asset which can be monetized by the entity's bond holders upon their own default (explained in more detail in Section 2. Whenever an entity's credit improves instead, it is effectively taxed as its DVA depreciates. Wealth is thus transfered from the equity holders of successful companies to the bond holders of failing ones, the transfer being mediated by banks acting as financial intermediaries and implementing the traditional CVA/DVA mechanics. 

Rewarding failing firms with a cash subsidy may be a practice of debatable merit as it skews competition. But rewarding failing firms with a DVA asset is without question suboptimal from an economic standpoint: the DVA asset they receive is paid in cash from their counterparties. However, it cannot be invested and can only be monetized by bond holders upon default. The portion that is not monetized by bond holders, ends up sitting as a CVA reserve on banks' balance sheets. Either way, these capitals are largely sterilized.

Margin lending structures reverse the macro-economic vicious circles engendered by CVA/DVA mechanics by effectively eliminating long term counterparty credit risk insurance and avoiding the wealth transfer that benefits the bond holders of defaulted entities. This has several effects: (i) it accelerates the default of failing firms by extracting cash payments proportional to their credit spread and the fair value of the positive derivative exposure, (ii) it reduces recovery rates without damaging derivative counterparties but only bond holders of the defaulted entities, (iii) it gives an incentive to failing firms to orderly unwind their derivative position as they approach default, (iv) it reduces the carrying costs for derivatives trades to those more successful firms whose spreads tighten and gives them a competitive advantage by strengthening their ability to hedge. 

The impact of monetary and government policy in times of crisis is also radically different. Prior to the crisis, banks used to set aside a CVA reserve which is just the expected level of loss due to counterparty credit risk. Losses in excess of the CVA were not budgeted for and fell back to the banks' treasury departments. When the crisis arrived, the firms whose credit deteriorated had DVA claims that overwhelmed banks and forced governments to intervene with major cash injections. These funds selectively benefited failing firms whose DVA gains were the highest. High quality names with relatively tighter credit spreads instead did not draw any benefit. The government policy of providing subsidies in terms of DVA financing perhaps inadvertently favoured only the bond holders of defaulted entities who monetized the subsidy at the time of recovery. Perhaps unsurprisingly, this policy did not prove to be a particularly efficient allocation of resources and was unable to help reverting the economic recessionary phase. 

The Basel III Accord improves on the pre-crisis situation by allowing for CVA VaR charges which effectively raise the bar for capital adequacy and strengthen provisioning strategies. More importantly, the Basel III Accord also offers the alternative of full collateralization which solves the problem of counterparty credit risk at the root (except in cases of extreme contagion), thus eliminating the anti-economical DVA subsidies to the bond holders of bankrupt firms. However, full collateralization raises the spectrum of heightened liquidity risk unless a robust and extensive market infrastructure for margin lending is established. 

Assuming an infrastructure for margin lending extensive enough to cover the entire scope of derivative markets, a systemic financial crisis could still be caused by a mismatch between offer and demand in the hypothecs market. This would happen if margin lenders cannot provision sufficient hypothec capital buffers from investors to lend to participants in derivative markets so that they can meet their collateral posting obligations in full. The existence of a fully developed market infrastructure for margin lending would help averting a systemic crisis by providing a new tool for monetary authorities to inject liquidity temporarily in the form of short-term, super-senior hypothec financing to margin lenders. This sort of intervention would still be an exceptional occurrence, but would arguably be less intrusive from a macro-economic standpoint than quantitative easing and the heavy injection of government funding into the banking system. While a crisis could be mitigated by granting super-senior hypothecs, it would still lead to a widening of credit spreads and extract wealth from failing firms in the form of insurance premia, thus accelerating or even triggering defaults and ultimately reducing recovery rates on defaulted debt, to the detriment of debt holders. 

We foresee margin lenders in OTC markets to be financial intermediaries negotiating hypothecs. While on the asset side there are insurance wraps on netting sets created through full collateralization by hypothecs to qualifying collateral, on the liability side on the balance sheet of margin lenders there are investors providing transferable hypothecs to qualifying collateral of which they retain ownership. The hypothec notes issued by margin lenders will arguably be ranked by seniority and be similar to a securitization structure intermediate between a cash and synthetic CDO. Similarly to synthetic CDOs, hypothec tranche payments do not include riskless interest rates but are only based on credit spreads as the beneficial owners of the assets underlying hypothecs are receiving interest payments independently. However, similarly to a cash CDO, hypothecs are fully collateralized and unlevered. Unlike both traditional synthetic and cash CDOs, we envisage hypothecs to be short term instruments with maturities of about 6 months as their objective is to ensure that CVA volatility risk is not transferred as with standard CVA/DVA structuring, but rests with the originating counterparty.

The CDO market played a trigger role in the crisis started in 2007 in part because of the inadequate, over-simplistic and non-rigorous local valuation methodologies such as the one-factor Gaussian copula model which did not take advantage of modern computing technologies, see for instance \cite{BrigoPallaTorreWiley}. The insistence on bipartite CVA structuring with the consequent large scale wealth transfer and anti-economical misallocations are similarly not due to a conscious policy decision. Instead, they derive from the use of inadequate and simplistic methodologies that are not technologically current and only allow the valuation of linear risk metrics which are indifferent to credit-credit correlations across counterparties such as the traditional forms of CVA.  Moving toward margin lending will thus require a wave of renewal and a realignment between mathematical modeling and computing technologies along with rigorous and consistent global valuation methodologies. 
 
\subsection{Organization of the paper}

The paper is organized as follows. Section \ref{definitions_and_axioms} contains definitions and axioms that acceptable counterparty credit risk structures need to satisfy. Section \ref{sec_inconsistent_structures_for_vulnerable_transactions} reviews the three historical standards for counterparty credit risk that dominated in the periods prior to 2007, between 2007 and 2009 and post 2009. All three standards are inconsistent from the accounting and valuation standpoint. As we explain, inconsistencies have gradually been recognized and reduced through time but were not entirely eliminated. In Section \ref{sec_first_to_default_bilateral_cva}, we discuss the first-to-default CVA, the first fully consistent structure that has been proposed. In Section \ref{sec_portable_cva}, we introduce two alternative forms of portable CVA. In Section \ref{sec_margin_lending}, we discuss collateralized transactions based on a market infrastructure for margin lending. Each of the Sections \ref{sec_inconsistent_structures_for_vulnerable_transactions}, \ref{sec_first_to_default_bilateral_cva}, \ref{sec_portable_cva} and \ref{sec_margin_lending} includes a comparison in terms of consistency from an accounting and valuation standpoint, hedgeability, numerical complexity, portability upon default, induced behaviour and macro-economic impact.
In Section \ref{penta_partite}, we discuss dynamic replication and in Section \ref{finite_liquidity} we consider the case of finite liquidity. Section \ref{conclusions} concludes.

Finally, we would like to point out that this is not meant to be a fully rigorous mathematical paper. We believe the mathematics to be sound but we have not formalized it fully as our objective is to provide an informal introduction. We will formalize the framework in full technical and rigorous detail in future work.

\section{Definitions and axioms}
\label{definitions_and_axioms}

For the probabilistic setup and our use of arbitrage freedom we refer to the appendix. 

Consider two defaultable entities $B$ and $C$, exchanging stochastic cash flows from time $0$ and up to the future final time $T$.
Let $\tau_B$ and $\tau_C$ be the stopping times for default of $B$ and $C$ respectively. Also, let
$R_t(B)$ and $R_t(C)$ be the recovery rates in case $B$ and $C$ default at time $t$, respectively. Finally, let $r_t$ be the risk-free rate and let's agree to work under the risk neutral measure for pricing. 

According to the Fundamental Theorem of Finance,  \cite{BDF1931}, \cite{DelbaenSchachermayer}, \cite{HarrisonPliska},  there is no arbitrage if and only if all market transactions between any pair of non-defaultable entities, i.\,e., entities where $\tau_B = \tau_C = \infty$ with probability one,
are valued as the discounted expectation of future cash flows under a globally defined valuation measure. Therefore, in the default-free case, we denote the related fair value to $B$ by $M_t(B)=- M_t(C)$. We assume this fair value has been obtained through the risk neutral valuation formula as an expectation of future discounted cash flows under one and the same globally calibrated measure used across all instruments (cf. e.g. \cite{BieleckiRutkowski2002}, Definition 8.1.2.). We also assume that there is a final maturity $T < \infty$ such that no cash flows occur after time $T$, i.\,e., $M_t(B)= M_t(C) = 0 \;\forall t>T$

The process $r_t$ is the short-term interest rate and is assumed to be adapted to the market filtration at time $t$. Similarly, ${\Bbb{E}}_t$ denotes expectation conditional on the same filtration\footnote{Mathematical modeling of filtration and processes will be made explicit in a subsequent paper.}. 

Let $V_t(B)$ and $V_t(C)$ denote the fair values to parties $B$ and $C$ respectively of the transaction. Defaultable transactions are not directly covered by the Fundamental Theorem since the life of a valuation agent is limited by its own default arrival time. 

In \cite{Brigo2005} the case of the CDS options market model is analyzed, where equivalence of measures is restored through a complicated approach based on a restricted filtration which takes out the default singularity when taking expectations, see also \cite{BrigoMorini2010} for the multi-name version of the same technique.  

We can nevertheless avoid such technicalities and reduce the problem of fair valuation to the Fundamental Theorem by decomposing a defaultable transaction in terms of an economically equivalent portfolio of hypothetical transactions exchanged between two hypothetical non-defaultable entities and which explicitly includes default protection contracts on $B$ and $C$. This is similar, in the bipartite case, to $B$ and $C$ buying protection through contingent CDS from default-free parties to perfectly hedge the respective unilateral CVA's with respect to C and B. 

\begin{definition}{\it

In case at least one of the two parties $B$ or $C$ is defaultable and there are no collateral posting provisions ensuring a 100\% safe level of overcollateralization, i.\,e., if the recovery rate on the fair value of a derivative transaction is not one by construction, then the transaction is called {\bf uncollateralized}. If $B$ and $C$ are obligated to post full collateral as a guarantee through a margin lender with a safe buffer of overcollateralization that is safe with probability 1, the transaction is {\bf collateralized}. We use this extreme notion of collateralization to simplify exposition but the assumption can be relaxed. 

If only two parties $B$ and $C$ are involved and they exchange protection with each other, the transaction is called {\bf bipartite}. If in addition $C$ buys protection on its own default from a margin lender $A$ while still $B$ buys protection on its own default from $C$, the transaction is {\bf tri-partite}. If $B$ buys protection from a margin lender $D$ and $C$ buys protection from a margin lender $A$, the transaction is {\bf quadri-partite}. Finally, if there is a Central Counterparty (CCP) that stands as the counterparty to all trades as in Figure 2, the transaction is {\bf penta-partite}.

}

\end{definition}

\begin{rem}
In a traditional bipartite transaction one may be used to CVA and DVA terms being calculated in a standard way. Such terms are equivalent to the positions mentioned above in the following sense. 

The CVA term B charges to $C$ is given by the loss given default fraction of the residual exposure measured by $B$ at default of the counterparty $C$ if such exposure is positive. This is the loss faced by $B$ upon default of $C$. If $C$ does not pay such loss to $B$, this is like saying that $B$ is condoning such loss to C, or in other terms B is selling protection to C on the traded portfolio contingent on C's own default. This would be like B offering a contingent CDS to C, offering protection against default of C to C and referencing the traded portfolio between B and C. A similar analogy holds for the DVA term and C selling protection to B against B's own default.   
\end{rem}


We now move to analyse more in detail the axioms that CVA and DVA price processes should satisfy. In the following we are not referring to a particular formulation of CVA or DVA, unless explicitly mentioned. We are rather writing down the requirements that any sensible definition of CVA and DVA should satisfy.


\subsection{Uncollateralized bipartite transactions}

If the transaction is bipartite and uncollateralized, then party $B$ sells to party $C$ default protection on $C$ contingent to an amount specified by a {\bf close-out rule}. Vice versa on the default of $B$, party $C$ sells protection to $B$ on the default of $C$.

In formulas:
\begin{align}
V_t(B) &= M_{t}(B) - \CVA_t(B, C) + \DVA_t(B, C), \nonumber\\
V_t(C) &= M_{t}(C) - \CVA_t(C, B) + \DVA_t(C, B), \label{eq_v_t_c}
\end{align}
where
\begin{itemize}
\item{} $M_{t}(B)$ is the mark to market to $B$ in case both $B$ and $C$ are default-free;
\item{} $\CVA_t(B, C)$ is the value of default protection that $B$ sells to $C$ contingent on the default of $C$, as assessed by $B$ at $t$.
\item{} $\DVA_t(B, C)$ is the value of default protection that $C$ sells to $B$ contingent on the default of $B$, as assessed by $B$ at $t$.
\item{} Similar definitions extend in case we exchange $B$ and $C$, i.\,e. $B \leftrightarrow C$.
\end{itemize}

The above two valuation formulas reflect the decomposition of the trade as a sum of a riskless derivative transaction and reciprocal default protection contracts. For all the specific definitions of CVA or DVA that we examine below, we require that the valuation is decomposed as in the both equations \ref{eq_v_t_c}. 

It will occasionally be helpful to write \ref{eq_v_t_c} exactly at a default time. For instance, in the scenarios where $\tau_B < \tau_C$, we write
\begin{equation}\label{eq:dva-cvaproof} \DVA_{\tau_B}(C, B) - \CVA_{\tau_B}(C, B) = V_{\tau_B}(C) - M_{\tau_B}(C) 
\end{equation}

In order for the valuation of an uncollateralized, bipartite transaction to be fully consistent, the axioms $A_1$, $B_1$ below must be valid along with one of the close-out conditions below, either $C_1$ or $C_2$:
{\it
\begin{itemize}
\item[($A_1$)] { Discounted martingale condition until default}:
The process $\CVA_t(B, C)$ is defined for all $t \leq \tau_B\wedge\tau_C$ and satisfies the equation
\begin{equation}
\CVA_t(B, C) = {\Bbb{E}}_t\left[ e^{-\int_t^{\tau_B\wedge\tau_C} r_s ds} \CVA_{\tau_B\wedge\tau_C}(B, C)  \right].
\end{equation}
\item[($B_1$)] {Money conservation until default}:
\begin{align}
\CVA_t(B, C) = \DVA_t(C, B),\\
\DVA_t(B, C) = \CVA_t(C, B),
\end{align}
for all $t \leq \tau_B\wedge\tau_C$.
\item[($C_1$)] {\bf Risk-free close-out rule:} If $\tau_B<\tau_C$, then
\begin{align}
V_{\tau_B}(C) &= -(M_{\tau_B}(C))^- + R_{\tau_B}(B) (M_{\tau_B}(C))^+ = M_{\tau_B}(C) - (1-R_{\tau_B}(B))(M_{\tau_B}(C))^+\,.
\label{eq_c1_0}\end{align}
Here $a^+ : = \max\{a, 0\}$, $a^- : = a^+ - a = (-a)^+$ for all reals $a$. $V_{\tau_B}(C) $ is interpreted as the value to $C$ of the transaction at the time when $B$ defaults. Consequently, due to \ref{eq:dva-cvaproof} we can recast equation \ref{eq_c1_0} as follows:
\begin{align}
\CVA_{\tau_B}(C, B) = (1 - R_{\tau_B}(B)) (M_{\tau_B}(C))^+ + \DVA_{\tau_B}(C, B).
\label{eq_c1}
\end{align}
Similar conditions with $B \leftrightarrow C$ also hold.
\item[($C_2$)] {\bf Replacement close-out rule:} If $\tau_B<\tau_C$, then
\begin{align}
V_{\tau_B}(C) &= - (M_{\tau_B}(C) + DVA_{\tau_B}(C, B))^- + R_{\tau_B}(B) (M_{\tau_B}(C) + \DVA_{\tau_B}(C, B))^+.\label{eq_c2_0}\\
&= M_{\tau_B}(C) + DVA_{\tau_B}(C, B) - (1-R_{\tau_B}(B))(M_{\tau_B}(C) + DVA_{\tau_B}(C, B))^+\nonumber \,.
\end{align}
Equivalently $($due to \ref{eq:dva-cvaproof}$)$ we have that
\begin{align}
\CVA_{\tau_B}(C, B) = (1 - R_{\tau_B}(B)) (M_{\tau_B}(C) + DVA_{\tau_B}(C, B))^+.
\label{eq_c2}
\end{align}
Similar conditions with $B \leftrightarrow C$ also hold.
\end{itemize}
}

In condition $C_2$, notice that the amount $(1 - R_{\tau_B}(B)) (M_{\tau_B}(C) + \DVA_{\tau_B}(C, B))^+$ can be interpreted as the novation cost, i.\,e., the loss amount to $C$ deriving from the default of $B$ and assuming that the same derivative transaction is novated with a default-free counterparty at time $\tau_B$. A similar financially motivated interpretation is not available in the case of a risk-free close-out as in $C_1$ as this rule does not reflect correctly the novation cost to the surviving party.

\subsection{Collateralized quadri-partite transactions with high frequency resets}

Collateralized transactions are best interpreted as transactions which reset periodically in time whereby $C$ buys default insurance from a margin lender $A$ and $B$ buys default insurance from a margin lender $D$. See Figure 1.

The case of high frequency time resets on short periods $\Delta t$ is mathematically the simplest, although not quite realistic as very frequent credit spread resets would imply major liquidity risk. We assume here that there is no jump in the risk free valuation $M$ of the contract when default happens. This is not the case for Credit Default Swaps as underlying instruments under a variety of models, see for example \cite{BrigoCapponiPallavicini}. In our case, though, \textit{we assume there are no jumps and there is no instantaneous contagion} and as a consequence the contractual structures are designed in such a way to 
periodically reset valuations with (equidistant) period $\Delta t > 0$
at the following equilibrium levels ($i = 0, 1, \ldots , n = \frac{T}{\Delta t}\}$):
\begin{align}
V_{i \Delta t}(A) &=  0 \\
V_{i \Delta t}(B) &= M_{i \Delta t}(B),\\
V_{i \Delta t}(C) &= M_{i \Delta t}(C),\\
V_{i \Delta t}(D) &=  0.
\label{eq_repo}
\end{align}
To achieve this objective
the margin lender $A$ secures in a segregated custodian account a sufficient amount of collateral to guarantee that the amount $(M_{t}(C))^-$ is paid to $B$ in case $C$ defaults at $t$, thus offsetting the counterparty credit risk that $C$ would otherwise pose to $B$. In return, $C$ pays to $A$ a stream of premia payments $\Delta \Pi_t(A, C)$ with period $\Delta t$. Symmetrically, $B$ pays a cash flow stream $\Delta \Pi_t(D, B)$ to $D$ so to ensure that $C$ is immunized from the risk of default of $B$.

In the case of this structure, the arbitrage free valuation axiom reads as follows:
{\it
\begin{itemize}
\item[($A_2$)] { Discounted martingale condition until default}:
\begin{align}
&{\Bbb{E}}_t[\Delta V_t(A) + 1_{t < \tau_C < t+\Delta t} (1-R_{{\tau_C}}(C))  (M_{\tau_C}(C))^-  + \Delta  \Pi_{t}(A, C)] = 0, \\
&{\Bbb{E}}_t[\Delta V_t(D) + 1_{t < \tau_B < t+\Delta t} (1-R_{{\tau_B}}(B))  (M_{{\tau_C}}(B))^-  + \Delta  \Pi_{t}(D, B)] = 0.
\end{align}
for all $t<\tau_B\wedge\tau_C$. Here, $\Delta V_t(A) \equiv V_{t+\Delta t}(A) - V_t(A)$ and we neglect discounting as is legitimate in the asymptotic limit $\Delta t\to 0$.
\end{itemize}
}

\begin{figure}[t]
\begin{center}
\includegraphics[width = 11cm]{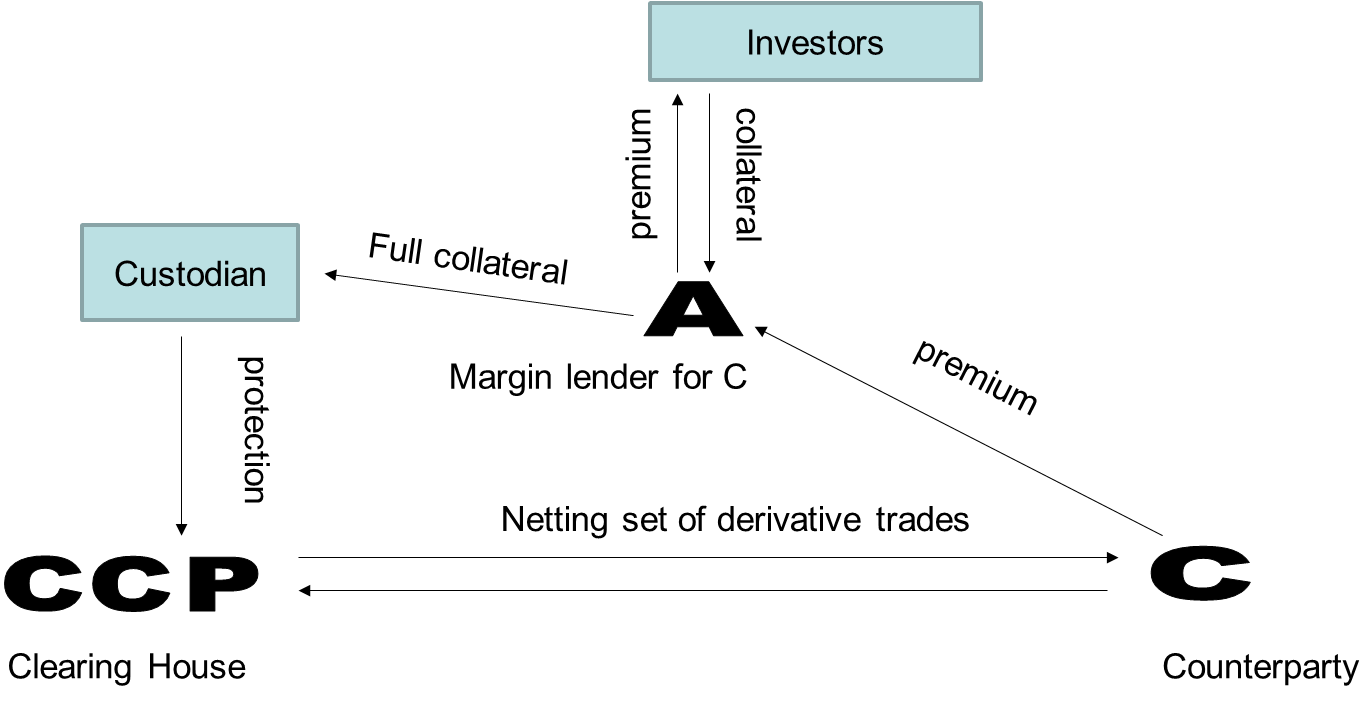}
\end{center}
\caption{Tri-partite scheme whereby the counterparty $C$ posts collateral to the bank $B$ by borrowing the collateral from the margin lender $A$.}\label{fig_tri}
\end{figure}

\subsection{Collateralized tri-partite transactions with periodic resets}

The tri-partite case is described in Fig. \ref{fig_tri} and correponds to a situation where the counterparty $C$ has the obligation to post collateral in full to the bank $B$, while the bank $B$ does not have a similar obligation. 

If resets are periodic and occur at times $T_i, i = 0, 1, 2, \ldots , n$, the fair value of the position to the margin lenders is zero only at the reset dates $T_i$. However, the margin lender $A$ sells default protection to $C$, a contract whose value is $\CVA_{t}(A, C)$. Similarly, $D$ sells default protection to $B$.
\begin{align}
V_{t}(A) &= e^{\int_{T_i}^t r_s ds} \Pi_{T_i}(A, C) - \CVA_{t}(A, C),\\
V_{t}(B) &= M_{t}(B) + \DVA_{t}(B, C),\\
V_{t}(C) &= M_{t}(C) - \CVA_{t}(C, B)  + \DVA_{t}(C, A),\\
\end{align}

{\it
\begin{itemize}
\item[($A_3$)] {Discounted martingale conditions until default}: For all $t\in [T_i, T_{i+1}]$, the CVA terms satisfy the equation
\begin{align}
\CVA_{t}(A, C) &= {\Bbb{E}}_t[e^{-\int_t^{\tau_C} r_s ds} 1_{\tau_C < \tau_B} 1_{\tau_C < T_{i+1}} (1-R_{\tau_C}(C))  (M_{\tau_C}(C))^-], \hskip1cm \forall t<\tau_C\\
\CVA_{t}(C, B) &= {\Bbb{E}}_t[e^{-\int_t^{\tau_B} r_s ds} (1-R_{\tau_B}(B))  (M_{\tau_B}(B))^-], \hskip1cm \forall t<\tau_B
\end{align}
and the premium received by $A$ from $C$ is computed so that
\begin{align}
\Pi_{T_i}(A, C) = \CVA_{T_i}(A, C),
\end{align}
\item[($B_1$)] { Money conservation until default}:
\begin{align}
\CVA_t(A, C) = \DVA_t(C, A),\\
\CVA_t(C, B) = \DVA_t(B, C),
\end{align}
for all $t<\tau_B\wedge\tau_C$,
\end{itemize}
}

\subsection{Collateralized quadri-partite and penta-partite transactions with periodic resets}

The quadri-partite case is described in Fig. \ref{fig_abcd} and correponds to a situation where both, the counterparty $C$ and the bank $B$ have a contractual obligation to post collateral in full to each other.

In the case of periodic resets at times $T_i, i = 0, 1, 2, \ldots, n$, the fair value of the position to the margin lenders is zero only at the reset dates $T_i$. However, the margin lender $A$ sells default protection to $C$, a contract whose value is $CV\!A_{t}(A, C)$ at $t$. In return, $A$ obtains from $C$ premium
payments $\Pi_{T_i}(A, C)$ at times $T_i, i = 0, 1, \ldots , n$.  Similarly, $D$ sells default protection to $B$ and receives periodic premium payments $\Pi_{T_i}(D, B)$ at $T_i, i = 0, 1, \ldots , n$. Assuming $t\in [T_i, T_{i+1}]$, 
\begin{align}
V_{t}(A) &= e^{\int_{T_i}^t r_s ds} \Pi_{T_i}(A, C) - \CVA_{t}(A, C),\\
V_{t}(B) &= M_{t}(B) + \DVA_{t}(B, D),\\
V_{t}(C) &= M_{t}(C) + \DVA_{t}(C, A),\\
V_{t}(D) &= e^{\int_{T_i}^t r_s ds} \Pi_{T_i}(D, B) - \CVA_{t}(D, B)
\end{align}

{\it
\begin{itemize}
\item[($A_4$)] { Discounted martingale conditions until default}: For all $t\in [T_i, T_{i+1}]$ the CVA  terms satisfy the equation
\begin{align}
\CVA_{t}(A, C) &= {\Bbb{E}}_t[e^{-\int_t^{\tau_C} r_s ds} 1_{\tau_C < \tau_B} 1_{\tau_C < T_{i+1}} (1-R_{\tau_C}(C))  (M_{\tau_C}(C))^-],  \hskip1cm \forall t<\tau_C\\
\CVA_{t}(D, B) &= {\Bbb{E}}_t[e^{-\int_t^{\tau_B} r_s ds} 1_{\tau_B < \tau_C} 1_{\tau_B < T_{i+1}}  (1-R_{\tau_B}(B))  (M_{\tau_B}(B))^-], \hskip1cm \forall t<\tau_B
\end{align}
and the premia are computed so that
\begin{align}
\Pi_{T_i}(A, C) = \CVA_{T_i}(A, C), \\
\Pi_{T_i}(D, B) = \CVA_{T_i}(D, B)
\end{align}
\item[($B_2$)] { Money conservation until default}:
\begin{align}
\CVA_t(A, C) = \DVA_t(C, A),\\
\CVA_t(D, B) = \DVA_t(B, D),
\end{align}
for all $t<\tau_B\wedge\tau_C$.
\end{itemize}
}

The penta-partite case where a CCP provides a clearing service is described in Fig. \ref{fig_ccp}. If the CCP accepts segregated collateral posted at custodian accounts for variation margin this case is economically equivalent to the quadri-partite case above. In case the CCP insists on receiving variation margin in cash instead (as is the case for instance with the LCH) then one requires an additional intermediary that provides a collateral transformation service, i.\,e., issues a revolving line of credit backed by the segregated collateral.

\section{Inconsistent structures for uncollateralized transactions}
\label{sec_inconsistent_structures_for_vulnerable_transactions}

Prior to FAS 156-157, standard structuring of counterparty credit revolved around inconsistent formulas that did not satisfy the conditions in the previous section. The standard was to consider the unilateral CVA defined as follows:
\begin{align}
\UCVA_t(B, C) &=  {\Bbb{E}}_t\left[ e^{ -\int_t^{\tau_C} r_s ds}  (M_{\tau_C}(C))^-  (1-R_{\tau_C}(C))  \right], \hskip1cm \forall t < \tau_C
\end{align}
while the UDVA term was set to zero.
In this case, the discounted martingale condition $A_1$ is satisfied by construction. The risk-free close-out rule $C_1$ is also satisfied since, in case $\tau_B<\tau_C$, we have that
\begin{equation}
\UCVA_{\tau_B}(C, B) = (1 - R_{\tau_B}(B)) (M_{\tau_B}(C))^+
\label{eq_CVA_2007}
\end{equation}
and $\UDVA_{\tau_B}(C, B) = 0$. By the same token, also condition $C_2$ is trivially satisfied as the UDVA is zero. However, the money conservation condition $B_1$ is not satisfied in this case.

After FAS 156-157, the standard changed to include the UDVA, i.\,e.
\begin{align}
\UDVA_t(B, C) &= \UCVA_t(C, B)
\end{align}
and accordingly $B \leftrightarrow C$. This modified definition satisfies condition $B_1$ but compromises the validity of $C_1$ and $C_2$ as the DVA term is now non-zero.

Notice that the close-out condition $C_1$ fails because of a mismatch equal to $\UDVA_{\tau_B}(C, B)$ between the right hand sides of equations \ref{eq_CVA_2007} and equation \ref{eq_c1}.

In 2009, the ISDA introduced the replacement close-out condition in $C_2$ with the intent to replace the risk-free close-out condition $C_1$ that prevailed until then, thus reducing the mismatch to
\begin{equation}
(1 - R_{\tau_B}(B)) \big((M_{\tau_B}(C) + \UDVA_{\tau_B}(C, B))^+ - (M_{\tau_B}(C))^+ \big).
\end{equation}
Notice however that this mismatch is still non-zero and the accounting inconsistency was not entirely eliminated: the impact of the loss of DVA upon default was reduced by a factor equal to the loss-given-default $1-R_{\tau_B}(B)$, but still this loss was not properly valued and accounted for.

Besides all being plagued by one inconsistency or another, these three heritage structures have the following characteristics:
\newpage
\begin{itemize}
\item[(i)] { \it Hedgeability:}
\begin{itemize}
\item The UDVA terms are problematic as hedging their variation would involve a party selling protection on itself, an impossible feat. The practice of hedging by proxying, namely selling protection on a name that is strongly correlated, can partly reduce spread risk but exacerbates jump to default risk and systemic risk. Indeed, if the name on which protection is sold actually defaults, the seller who was hedging her DVA needs to make the protection payment, and this could push the seller herself into a worse credit situation and closer to default. For example, back in 2008, a hypothetical top investment bank that had been highly correlated with Lehman Brothers and who had decided to hedge her DVA by selling protection on Lehman would have been in deep troubles. For informal comments on DVA hedging see for example \cite{BrigoFAQ}. 
\item The UCVA variability instead can in principle be hedged, assuming the ability to accurately value sensitivities and to assess gap risk. For an example on the assessment of the pricing component of gap risk see for example \cite{BrigoCapponiPallavicini}.
\end{itemize}
 
\item[(ii)] {\it  Numerical complexity:} Valuing the UCVA involves modeling only one credit dynamically along with all market factors affecting the derivative portfolio. The UCVA is not sensitive to credit correlations, unless the underlying portfolio contains credit instruments. In particular, UCVA is not sensitive to default correlation between $B$ and $C$. However, the UCVA is sensitive to credit-market correlations.
\item[(iii)] {\it  Portability:} The structures with replacement close-out are relatively more portable than the structures with risk-free close-out. However, since there is a mismatch in condition $C_2$ and the loss of UDVA is not correctly priced and hedged dynamically, a novation transaction in case of default involves a net unhedged loss for the surviving party.
\item[(iv)] {\it  Induced behaviour:} As the credit of one party deteriorates, its UDVA raises and the party can realize a gain. On the other hand, as the credit of one party improves, its UDVA lowers and the party needs to realize a balance sheet loss. This effect rewards credit degradation and is an incentive on banks not to have a top credit rating as achieving a high rating would force to realize accounting losses.
\item[(v)] {\it  Macro economic impact:} The forced mutual sale of default protection embedded in all derivative transactions, implicit in the CVA and DVA practices, is a mechanism to transfer wealth from the equity holders of profitable companies with good credit to the bond holders of companies with bad credit. This is a consequence of the earlier point (iv). 
\item[(vi)] {\it  Bank capitalization:} Substantial bank capital needs to be allocated to UCVA reserves. In absence of collateral capital also needs to be allocated to compensate for UCVA volatility which is reflected in the UCVA VaR capital charge in Basel III.  Finally, there is a third charge against default risk due to the presence of substantial gap risk and hedging inefficiencies, as we hinted at above.

 By definition, the UCVA is the expected loss due to defaults and the loss distribution is highly skewed and very fat tailed. Hence, a UCVA provisioning strategy gives risk to systematic small gains for the UCVA desk and occasional large losses which are apportioned unexpectedly to banks' treasury departments.
\end{itemize}

\section{First-to-default CVA}
\label{sec_first_to_default_bilateral_cva}

First-to-default clauses appear in \cite{DuffieHuang}, \cite{BieleckiRutkowski2002} and are made explicit for CVA and DVA calculations in \cite{BrigoCapponi2010}, \cite{BrigoCapponiPallavicini}, \cite{BrigoCapponiPallaviciniPapatheodorou}, \cite{BrigoPallaviciniPapatheodorou}, \cite{BrigoMorini2011}, see also \cite{Gregory2009}.

In the first-to-default bilateral structure, the CVA and DVA are defined as follows:
\begin{align}
\FTDCVA_t(B, C) &= \FTDDVA_t(C, B) = {\Bbb{E}}_t\left[ 1_{\tau_C<\tau_B} e^{ -\int_t^{\tau_C} r_s ds}  (M_{\tau_C}(C))^-  (1-R_{\tau_C}(C))  \right], \hskip0.1cm \forall t < \tau_C\,.
\end{align}
and $B\leftrightarrow C$.
If $\tau_B<\tau_C$, we have that $\FTDCVA_{\tau_B}(B, C) = \FTDDVA_{\tau_B}(C, B) = 0$. Hence, conditions $C_1$ and $C_2$ are actually equivalent in this case and they are both satisfied. (To avoid confusion this claim is not in contradiction with what is stated in \cite{BrigoMorini2011}: in \cite{BrigoMorini2011}, the first-to-default bilateral CVA is shown to be inconsistent under a replacement closeout rule which is different from the rule we consider in this paper, in that the payout is expressed in terms of unilateral DVA).

The FTDCVA structure is consistent from a valuation standpoint. Other characteristics include the following:
\begin{itemize}
\item[(i)] {\it  Hedgeability:}  Unlike the unilateral CVA in the previous section, the first-to-default CVA of one entity is sensitive to the entity's own distance to default and decreases with this distance tending to zero in the limit where the entity itself defaults. Since the FTDCVA is typically much larger than the FTDDVA of a bank, the non-hedgeability issue is potentially highly problematic, more so than the already worrisome unhedgeability of the UDVA in the unilateral case.
\item[(ii)] {\it  Numerical complexity:} The FTDCVA is sensitive to credit correlations between $B$ and $C$ as these have an impact on the first-to-default clause. Hence a fully correlated credit-market simulation is necessary.
\item[(iii)] {\it  Portability:} The structures are less portable than in the unilateral case with replacement close-out rules as the first-to-default DVA is null in case the default of one of the two entities occurs. Hence, a novation against a non-defaulted counterparty entails a loss equal to the DVA of the surviving entity.
\item[(iv)] {\it  Induced behaviour:} As the credit of one party deteriorates and approaches default while the credit of the counterparty stays stable its FTDCVA goes to zero but not its FTDDVA. Hence, the valuation of a derivative transaction raises and ends up above the risk free level. In particular, entities approaching default benefit from entering derivative transactions.
\item[(v)] {\it  Macro economic impact:} The wealth transfer from the equity holders of successful companies to the bond holders of defaulted entities is more pronounced than in the standard inconsistent structures in the previous section because of the wrong-sign sensitivity of the FTDCVA to the credit of the assessing entity.
\item[(vi)] {\it  Bank capitalization:} Currently, the Basel III Accord insists on using the unilateral CVA as a metric to determine capital requirements. The possibly large discrepancy between the first-to-default CVA and the unilateral CVA is problematic when the former is used for valuation purposes as it generates conflicting priorities for risk management. Moreover, the FTDCVA collected from clients is not sufficient to provision UCVA capital. 

In a hypothetical scenario where the regulators were to endorse the FTDCVA for regulatory purposes, banks with higher credit worthiness would be forced to allocate more capital than banks with lower credit worthiness and would be less competitive. In the limit of a bank approaching default capital allocation requirements would tend to nil. Finally, as one goes through the credit cycle bank reserve requirements would be anti-cyclical, i.\,e., lower during recessionary periods in a higher credit spread environment and higher otherwise.
\end{itemize}

\section{Portable CVA}
\label{sec_portable_cva}

The unilateral process $UCVA(B, C)$ does not provide a consistent definition of CVA as it does not take into account the closeout conditions under neither $C_1$ nor $C_2$. A way to avoid this inconsistency is to resort to full fledged bilateral CVA, i.\,e., the first-to-default CVA we have seen earlier. However, we may also eliminate the inconsistency with a less dramatic change than going from UCVA to FTDCVA, as follows.

Let us start with the risk free closeout $C_1$. How can we depart as little as possible from unilateral CVA while having condition $C_1$ satisfied? Consider equation \ref{eq_c1} for $C_1$, and notice that it can be rewritten in terms of unilateral CVA, in scenarios where $\tau_B < \tau_C$, as 

\[ \CVA_{\tau_B}(C, B) = \mbox{UCVA}_{\tau_B}(C, B) + \DVA_{\tau_B}(C, B). \]

This holds at time $\tau_B$. If we try and write the analogous equation at time $t$,  then we obtain the following definition of CVA, which implies that party $B$ subtracts from its fair value calculation the discounted value of the DVA of $C$ that $B$ would be liable to pay upon defaulting, in case $B$ defaults prior to $C$. We thus introduce the portable CVA as the following process:
\begin{align}
\PCVA_t(C, B) : = \UCVA_t(C, B) + \Gamma_t(C, B), \hskip1cm \forall t < \tau_B
\label{eq:PCVAgamma}
\end{align}
where
\begin{equation}
\Gamma_t(C, B) : = {\Bbb{E}}_t\left[ 1_{\tau_B<\tau_C}  e^{ -\int_t^{\tau_B} r_u du}  \UDVA_{\tau_B}(C, B) \right] 
\end{equation}
to satisfy the risk-free close-out condition in $C_1$. The important difference with FTDCVA is that here we have a UCVA term where we do not check who defaults first. 

Under the replacement close-out, the reasoning is analogous.  
We start from equation \ref{eq_c2} and add and subtract $\mbox{UCVA}_{\tau_B}(C, B)$ from the right hand side. We obtain, in scenarios where $\tau_B < \tau_C$, 
\[ \CVA_{\tau_B}(C, B) = \mbox{UCVA}_{\tau_B}(C, B) + \big( \left(M_{\tau_B}(C) + \UDVA_{\tau_B}(C, B)\right)^+  - (M_{\tau_B}(C))^+ \big) (1-R_{\tau_B}(B)) .\]
At this point the most obvious version of this equation at time $t$ is obtained as in \ref{eq:PCVAgamma} where, this time, $\Gamma$ is defined as
\begin{align}
\Gamma_t(C, B) = {\Bbb{E}}_t\left[ 1_{\tau_B<\tau_C}  e^{ -\int_t^{\tau_B} r_u du}  \big( \left(M_{\tau_B}(C) + \UDVA_{\tau_B}(C, B)\right)^+  - (M_{\tau_B}(C))^+ \big) (1-R_{\tau_B}(B)) \right].
\end{align}
Again, the important difference with FTDCVA is that here we have a UCVA term where we do not check who defaults first. 
Similar equations apply to the case $B\leftrightarrow C$. 

One can double-check that conditions $C_1$ and $C_2$ are indeed satisfied because, in case $\tau_C<\tau_B$, we have $\Gamma_{\tau_B}(C, B) = 0$ in both cases, so that 
\begin{equation}
\PDVA_{\tau_B}(C, B) \equiv \PCVA_{\tau_B}(B, C)
= \UCVA_{\tau_B}(B, C) \equiv \UDVA_{\tau_B}(C, B).
\end{equation}

Notice that under both $C_1$ and $C_2$ we have that $\Gamma_t(C, B)\ge0$. Hence, \begin{equation}
\PCVA_t(B, C) \ge \UCVA_t(B, C)
\end{equation}
for all $t$, the equality being attained only at the stopping time $\tau_B$.


As a ballpark estimate of order of magnitude and neglecting correlations between credit and market risk, the difference $\Gamma_t(C, B)$ between the portable CVA and the unilateral CVA is equal to the product between UDVA and the probability of default in case the UDVA and the default arrival time are uncorrelated. Due to the triangular inequality, in general the difference $\Gamma_t(C, B)$ is smaller than this amount. This estimate indicates that the difference $\Gamma_t(C, B)$ is relatively immaterial in most circumstances.

To conclude, this structuring style has the following properties:
\begin{itemize}
\item[(i)] {\it  Hedgeability:} The portable CVA is larger than or equal to the unilateral CVA and it is equal to the unilateral CVA in two opposite limits: when the entity's credit-worthiness tends to infinity (i.\,e., in the non-defaultable limit) and when the entity defaults. Similarly to the first-to-default bilateral structure FTDCVA, the sensitivity with respect to the credit of the entity assessing the PCVA could have the wrong sign for hedgeability. However, this sensitivity is quantitatively much smaller as it derives only from the term  $\Gamma_t(C, B)$, whose order of magnitude is roughly given by the product between the $DVA_t(C, B)$ and the probability to default before time $t$.
\item[(ii)] {\it  Numerical complexity:} The exact evaluation of portable CVA is more complex than the valuation of UCVA as it involves a nested simulation. However, approximation schemes can be devised to avoid the nested simulation and marginalize the computational impact of moving from UCVA to PCVA.
\item[(iii)] {\it  Portability:} The structure is nearly as portable as the unilateral case with replacement close-out rules as the portable DVA is equal to the UDVA upon counterparty default and the PDVA of the surviving entity against a novation counterparty does not deviate much from the UDVA.
\item[(iv)] {\it  Induced behaviour:} The portable CVA and PDVA are slightly larger than the unilateral CVA and UDVA. The incentive to enter into derivative transactions as default nears that is caused by the FTDCVA structure is absent in this case.
\item[(v)] {\it  Macro economic impact:} Similar to the standard case, just marginally better because of the consistency from an accounting standpoint.
\item[(vi)] {\it  Bank capitalization:} Both the PCVA and the PCVA VaR capital charges are slightly more conservative than those corresponding to the unilateral CVA.  As a consequence, if a bank collects PCVA from a counterparty, then it can use these funds to fulfill in full the UCVA capital adequacy obligation. 
\end{itemize}

\section{Collateralized transactions and margin lending}
\label{sec_margin_lending}

Valuation formulas in the tri-partite and quadri-partite cases are not ambiguous as the ones for the uncollateralized structures above. In fact, the valuation formulas have already been given as part of the risk neutrality conditions in Section \ref{definitions_and_axioms}.

The margin lenders $A$ and $D$ can procure hypothecs on qualifying collateral and place it in segregated accounts. $A$ would typically lend to several counterparties $C$ and obtain funding from various investors in tranches ranked by seniority. A tranche is characterized by an attachment level $L$ and a hypothec amount $N$. If in a (small) time interval $\Delta t$ the transaction covered by the hypothec suffers losses in excess of $L$, the investor will sustain these losses up to a maximum amount of $N$. In the quadri-partite case the fair premium (``CVA tranche spread'') to which an investor of the lender $A$ is entitled is valued as follows:  

\begin{equation}
V(A, L, N) = \frac{{\Bbb{E}}\Big[\sum_{i \geq 0} e^{-\int_0^{T_{i+1}}r_s ds}\big(X_{A,[L, L+N]}(T_{i+1}) - X_{A,[L, L+N]}(T_{i})\big)\Big]}{{\Bbb{E}}\Big[\sum_{i \geq 0} e^{-\int_0^{T_{i+1}}r_s ds} (T_{i+1} - T_{i}) \big(N - X_{A,[L, L+N]}(T_{i+1})\big)\Big]}, 
\end{equation}
where  
\[
X_{A,[L, L+N]}(t) : = \Big(\min \Big\{\sum_{C\in A} (M_{\tau_C}(C))^- (1-R_{\tau_C}(C)) 1_{\tau_C \leq t}, L+N\Big\} - L \Big)^+,
\]
$0 \leq t \leq T$, and $C\in A$ denotes that the sum is over all the counterparties served by $A$.
A similar equation holds for $B\leftrightarrow C$ and $A\leftrightarrow D$.

In the tri-partite case where $C$ is insured as opposed to $B$ the formula is different as it includes the DVA of $B$:
\begin{equation}
V(A, L, N) = \frac{{\Bbb{E}}\Big[\sum_{i \geq 0} e^{-\int_0^{T_{i+1}}r_s ds}\big(Y_{A,[L, L+N]}(T_{i+1}) - Y_{A,[L, L+N]}(T_{i})\big)\Big]}{{\Bbb{E}}\Big[\sum_{i \geq 0} e^{-\int_0^{T_{i+1}}r_s ds} (T_{i+1} - T_{i}) \big(N - Y_{A,[L, L+N]}(T_{i+1})\big)\Big]}, 
\end{equation}
where
\[
Y_{A,[L, L+N]}(t) : = \Big(\min \Big\{\sum_{C\in A} (- M_{\tau_C}(C) + DVA_{\tau_C}(B, C))^+ (1-R_{\tau_C}(C)) 1_{\tau_C \leq t}, L+N\Big\} - L \Big)^+
\]
for all $0 \leq t \leq T$.

The comparison between the case of collateralized transactions with margin lending and the previous uncollateralized structures reveals striking differences. In the quadri-partite case we have that:
\begin{itemize}
\item[(i)] {\it  Hedgeability:} One should distinguish between intra-period hedging and inter-period hedging. On an inter-period basis, since collateralized transactions reset, $B$ and $C$ are short their own credit. As a consequence, $B$ and $C$ can hedge their inter-period volatility by buying protection on themselves. On an intra-period basis instead, $B$ and $C$ are long a protection contract on themselves and this sensitivity is not hedgeable. However, the intra-period risk can be reduced by reducing the length of the reset period itself.

The margin lenders $A$ and $D$ are exposed to default risk and to gap risk. They can hedge default risk by securitization, while gap risk requires model driven strategies to allocate sufficient capital buffers made up of hypothecs and emergency revolving lines of credit in case a buffer overshoot event occurs. In the uncollateralized case instead, the ability to securitize is severely limited by the necessity to manage CVA volatility.

With collateralized structuring, the market risk caused by changes in fair value adverse to $C$ stays with $C$. Also the risk originating from the volatility of the credit spread of $C$ stays with $C$. The same can be said about $B$. Only the default risk for $B$ and $C$ are transferred to the margin lenders $A$ and $D$.

In a scenario where $B$ is the structuring bank, $B$ is left with replication risk and the risk on its own spread volatility. The two sources of risk are quite separate and can be hedged individually. As we discuss in the next section, the spread volatility risk suffered by $B$ in the course of dynamic replication can actually be eliminated in a penta-partite structure whereby the replication portfolio is cleared within the same netting set of the replication target. In the case of uncollateralized structures instead, CVA hedging is best performed within the trading book, a task which is very complex, delicate and error prone.

\item[(ii)] {\it  Numerical complexity:} Securitization requires the joint simulation of credit and market factors and the calculation of cumulative loss distributions. This task is just marginally more complex than the calculation of first-loss FTDCVA.

\item[(iii)] {\it  Portability:} collateralized contracts are portable by construction.

\item[(iv)] {\it  Induced behaviour:} Since $B$ and $C$ are short their own credit, if their credit improves they are rewarded with lower insurance payments while if their credit degrades they are subject to higher insurance payments. In particular, in case their credit degrades or there is an adverse move in the fair value of the transaction, they have an incentive to unwind or otherwise reduce their exposure. This behaviour mitigates systemic risk.

\item[(v)] {\it  Macro economic impact:} As compared to the uncollateralized structures considered above, the collateralized ones avoid entirely the transfer of wealth from the equity holders of successful companies to the debt holders of defaulted companies, thus rewarding successful firms.

\item[(vi)] {\it  Bank capitalization:} The CVA and CVA VaR charges are avoided entirely because the credit based derivative exposures have been replaced by asset based hypothecs. Banks do not need to provision CVA capital and do not need to hedge the CVA and DVA. Bank capital requirements are reduced but their counterparty risk is also massively reduced. Systemic risk would only resurface in case of a breakdown in the margin lending market infrastructure, a situation that liquidity provision facilities could be arranged to prevent.

\end{itemize}

In the tri-partite case, the same remarks apply but with the following few exceptions and differences:

\begin{itemize}
\item[(i)] {\it  Hedgeability:} $B$ still has a long term DVA term that entails an unheadgeable credit exposure to the spread of $B$ itself.

\item[(ii)] {\it  Numerical complexity:} The same considerations in the quadri-partite case apply here also.

\item[(iii)] {\it  Portability:} If $C$ defaults, the contract is portable from $B$'s standpoint. If $B$ defaults, the contract is still portable as the DVA on the balance sheet of $C$ is short term and it is against the margin lender $A$, not $B$.

\item[(iv)] {\it  Induced behaviour:} $B$ would behave as in the case of unilateral CVA while $C$ would behave as in the case of quadri-partite margin lending.

\item[(v)] {\it  Macro economic impact:} If the derivative exposures of buy-side firms were fully collateralized by tri-partite margin lending agreements, there would still be no transfer of wealth from the equity holders of successful companies to the bond holders of defaulted ones. There would however be a transfer of wealth from successful banks to defaulted ones.

\item[(vi)] {\it  Bank capitalization:} The same considerations in the quadri-partite case apply here also.

\end{itemize}

\section{Derivative replication by defaultable counterparties}
\label{penta_partite}

An initial analysis of the problem of replication of derivative transactions under collateralization but without default risk and in a purely classical Black and Scholes framework has recently been considered in 
\cite{Pieterbarg2010}. The fundamental impact of collateralization on default risk and on CVA and DVA has been instead analyzed in the already cited \cite{BrigoCapponiPallaviciniPapatheodorou} and \cite{BrigoCapponiPallavicini}. These works look at CVA and DVA gap risk under several collateralization strategies, with or without re-hypothecation, as a function of the margining frequency, with wrong way risk and with possible instantaneous contagion. Minimum thresholds amounts and minimum transfer amounts are also considered. The fundamental funding implications in presence of default risk have been considered in \cite{MoriniPrampolini2011}, see also \cite{Castagna2011}.

In this Section, we address this question from the angle adopted in this paper, i.\,e., we analyse transactions at the structuring level, resolve the defaultability features by expressing them as embedded options and ultimately make use of the Fundamental Theorem of Finance. The Fundamental Theorem does not assume replicability but makes only the much weaker hypothesis of absence of arbitrage. However, in the particular case where replication is possible, the valuation does coincide with the cost of replication.

Let's consider a derivative transaction for which, in case it is struck between two non-defaultable entities, there exists a perfect replication strategy for it. Having made this hypothesis, we then suppose that the same transaction is actually struck between two entities which can possibly default, a bank $B$ and a client $C$, and we ask the question: can the counterparty risk agreement be structured in such a way that the bank has available a strategy of perfect replication up to the stopping time of first default? If so, what is the cost of replication?

If the counterparty credit risk structure is inconsistent as in the three cases in Section \ref{sec_inconsistent_structures_for_vulnerable_transactions}, then there are multiple prices for the same payoff and the Fundamental Theorem does not apply. Inconsistent structures are a non-starter as the possibility of replication itself is an ill posed question.

In the remaining 6 structures discussed above, a fairly priced derivative transaction between defaultable entities is replicated by a portfolio of transactions between non-defaultable entities. The portfolio includes mutual default protection contracts whose valuation is given by the CVA and DVA. Notice that from this viewpoint, the CVA and DVA are not empirical adjustments which invalidate the Fundamental Theorem and thus require an entirely new theoretical framework for Finance not based on the principle of arbitrage freedom. From the viewpoint we have adopted, the Fundamental Theorem still applies also to the defaultable case, just as long as the CVA and DVA are interpreted themselves as the fair value of financial contracts. Hence, replication is possible if and only if the entire portfolio including the risk-free transactions and the CVA and DVA protection contracts can be replicated by the defaultable entities.

A necessary condition for replication is that the portfolio containing the risk-free transaction and the CVA and DVA terms for the structurer $B$ can be replicated at least in principle by a non-defaultable entity. This condition is necessary but not sufficient since a defaultable entity has at its disposal a reduced set of replication strategies as compared to non-defaultable entities. Particularly limiting is the fact that, if $B$ is defaultable, then $B$ cannot sell credit protection on itself.

The impossibility for an entity to sell protection on itself is a fundamental obstruction to replicability that applies in particular to the bilateral first-to-default CVA structure in Section \ref{sec_first_to_default_bilateral_cva} and the portable CVA structure in Section \ref{sec_portable_cva}. In this case, due to the wrong sign of the dependency of both the CVA and the DVA on the credit of the structurer, the combined CVA-DVA term cannot be replicated by a defaultable structurer.

Similar considerations apply to the tri-partite structure in Section \ref{sec_margin_lending}, where the structurer $B$ is also affected by a DVA term he himself cannot hedge.

The quadri-partite structure  in Section  \ref{sec_margin_lending} does not have an issue with DVA unhedgeability. However, unless the replication target and the hedging instruments are covered under a single overarching netting agreement, collateral posting obligations do not net and give rise to non-zero carry costs to a defaultable replicating entity.

To resolve the netting issue, the structure we can envisage that is most appropriate for replication purposes is a penta-partite variation to the quadri-partite structure. See Figure 2 for an illustration. In the penta-partite case, the $CCP$ is a Central Counterparty and $B$ is a structuring bank that engages in the replication of an exotic derivative custom designed for a particular client $C$. In case a dynamic replication strategy for the given transaction between non-defaultable entities exists, then the same strategy is also available to $B$ at the condition that $B$ clears through the same $CCP$ both the exotic derivative transaction {\bf and} the replicating portfolio. In this case, the net value to $B$ of the exotic derivative and corresponding hedges is zero and so is the DVA and the collateral funding cost.

Notice that:
\begin{itemize}
\item{} $B$ can replicate dynamically only up to the time of its default or the maturity date of the transaction,
\item{} the replication cost to $B$ is precisely equal to the risk-free value of the transaction, even if $B$ is defaultable and its cost of funding has a non-vanishing spread,
\item{} if counterparty $C$ chooses not to replicate with positions within the same netting set, it has a non-vanishing carry cost for the derivative replication strategy due to collateral posting obligations.
\end{itemize}

In case there are hedge slippages, then $B$ may possibly need to post collateral and sustain the corresponding cost of carry. Furthermore, a CVA and DVA term will emerge on the balance sheet of $B$. The combination of these three effects results in net costs that $B$ needs to transfer to the client ex-ante and price it as part of the structuring fee. Once received, this portion of fee needs to be set aside and provisioned as a risk reserve without being reflected in the fair valuation of the transaction.

\section{The case of finite liquidity}
\label{finite_liquidity}

There are two situations that would lead to losses over and above the open mark to market value at time of default: the case where the fair value process has a jump at time of default and the case whereby at least some of the derivative transactions in a netting set are not liquid and there is an extended close-out period to novate transactions over which the fair value may change. In both cases, if $C$'s defaults prior to $B$ the transaction may be novated or unwound by $B$ at time $\tau_C$ not at the fair value $M_t(B)$ but at the higher level $M_t(B) + L_{\tau_C}(B)$ where $L_{\tau_C}(B)\ge 0$ is a random variable giving the price correction deriving from the novation process.

Under a risk-free close-out rule, if $M_{\tau_C}(B)>0$, $B$ recovers the amount $(1-R_{\tau_C} ) M_{\tau_C}(B)$, while if $M_{\tau_C}(B)<0$ then $B$ pays this amount to $C$. Either way, novation costs are non-negative, i.\,e., $L_{{\tau_C}}(B)\ge 0$. Also, under a replacement close-out, the novation cost is the same.

In the case of bipartite transactions, the close-out rules are modified as follows in the presence of liquidity corrections:
\begin{itemize}{\it
\item[($C_1'$)] {\bf Risk-free close-out rule:} If $\tau_B<\tau_C$, then
\begin{align}
V_{\tau_B}(C) &= -(M_{\tau_B}(C))^- + R_{\tau_B}(B) (M_{\tau_B}(C))^+ - L_{\tau_B}(C).
\end{align}
Here, $V_{\tau_B}(C) $ is interpreted as the value to $C$ of the transaction at the time when $B$ defaults. Equivalently, we can recast these equations as follows:
\begin{align}
\CVA_{\tau_B}(C, B) = (1 - R_{\tau_B}(B)) (M_{\tau_B}(C))^+ + \DVA_{\tau_B}(C, B) + L_{\tau_B}(C).
\label{eq_c1_liq}
\end{align}
Similar conditions with $B \leftrightarrow C$ also hold. Equivalence of the two formulas is obtained by 
\item[($C_2'$)] {\bf Replacement close-out rule:} If $\tau_B<\tau_C$, then
\begin{align}
V_{\tau_B}(C) &= - (M_{\tau_B}(C) + \DVA_{\tau_B}(C, B))^- + R_{\tau_B}(B) (M_{\tau_B}(C) + \DVA_{\tau_B}(C, B))^+ - L_{\tau_B}( C).
\end{align}
Equivalently, we have that
\begin{align}
\CVA_{\tau_B}(C, B) = (1 - R_{\tau_B}(B)) (M_{\tau_B}(C) + \DVA_{\tau_B}(C, B))^+ + L_{\tau_B}(C).
\label{eq_c2_liq}
\end{align}
Similar conditions with $B \leftrightarrow C$ also hold.
}
\end{itemize}

In multi-partite structures with margin lending, liquidation costs would typically be partially borne by the margin lender up to a pre-assigned haircut level $H_{\tau_C}(B)$, and partly by either the novating party $B$ or the CCP that is responsible for losses in excess of $M_{\tau_C}(B)  + H_{\tau_C}(B)$.  Losses in excess of the haircut engender a CVA and DVA term which tends to zero as the haircut level tends to infinity. To keep things simple, in the following formulas we assume that $H_{\tau_C}(B) = \infty$ and pretend the margin lender absorbs all the losses. Under this simplifying assumption, we have that:
\begin{itemize}{\it
\item[($A_2'$)] {Secured quadri-partite transactions with high frequency resets}:
\begin{align}
&{\Bbb{E}}_t[\Delta V_t(A) + 1_{t < \tau_C < t + \Delta t} (1-R_{\tau_C}(C))  (M_{\tau_C}(C)^-  + L_{\tau_C}(C) )  + \Delta \Pi_{\tau_C}(A, C)] = 0
\end{align}
and $(A, C)\leftrightarrow (D, B)$, for all $t<\tau_B\wedge\tau_C$;

\item[($A_4'$)] {Secured quadri-partite transactions with periodic resets}: For all $t\in [T_i, T_{i+1}]$, the CVA satisfies the equation
\begin{align}
\CVA_{t}(A, C) &= {\Bbb{E}}_t[e^{-\int_t^{\tau_C} r_s ds} 1_{\tau_C < \tau_B} 1_{\tau_C < T_{i+1}} (1-R_{\tau_C}(C))  (M_{\tau_C}(C)^-+ L_{\tau_C}(C))]
\label{eq_cva_margin_lending_discrete_liq}
\end{align}
and $(A, C)\leftrightarrow (D, B)$. Premia are still computed so that
\begin{align}
\Pi_{T_i}(A, C) = \CVA_{T_i}(A, C)
\end{align}
and $(A, C)\leftrightarrow (D, B)$.
}
\end{itemize}

Having made these modifications in the close-out conditions, the corrected CVA and DVA which account for liquidity gaps are defined accordingly as discounted martingales consistently with the applicable stopping time conditions, i.\,e., either \ref{eq_c1_liq}, or \ref {eq_c2_liq} or \ref{eq_cva_margin_lending_discrete_liq}.

As an example of how liquidity considerations may impact valuation, consider the case of two treasury bonds which, after a given valuation date entail precisely identical future cash flows. Suppose also that one of the two bonds is on-the-run and highly liquid while the other bond is off-the-run and illiquid. Suppose that $B$ and $C$ exchange a repurchase agreement on a long-short combination in these two bonds, i.\,e., $B$ has the obligation to repurchase the on-the-run bond from $C$ while $C$ has the obligation to repurchase the off-the-run bond from $B$.

The framework in this paper allows one to determine the fair value of the transaction to either $B$ or $C$ in accordance with the Fundamental Theorem of Finance, i.\,e., in an arbitrage free fashion. To do so, one needs to account for the cost of carry of transactions of finite liquidity. More specifically, when valueing a transaction one should also specify the time horizon over which the transaction will be held and add the expected carry cost. In case $C$ wishes to value a strategy whereby the transaction is held until maturity or until either $B$ or $C$ default, the net cost is
\begin{align}
{\Bbb{E}}_t\big[ \int_t^{\tau_B\wedge \tau_C}  {\Bbb{E}}_s\big[ 1_{s < \tau_C < s+ds} \big( (1-R_{\tau_C}(C))  L_{\tau_C}(B, C) \big) \big]  \big].
\end{align}
Hence, the strategy to holding a direct REPO in an off-the-run bond and an inverse REPO in an equivalent on-the-run bond has a positive arbitrage free cost to $C$.
The total expected cost to $C$ is zero if one includes the putative gain that $C$ would make in case of its own default. The opposite strategy has a zero cost to $B$ assuming that $C$ posts collateral in full through a margin lender. The strategy of the margin lender has zero fair value at equilibrium.

\section{Conclusions}
\label{conclusions}

We have reviewed 10 different structuring styles for counterparty credit risk sensitive transactions. In our analysis, we reduce to the Fundamental Theorem of Finance by interpreting the CVA and DVA terms as prices of embedded contractual obligations contingent to the event of default. As such, the various different definitions of CVA and DVA are precisely identified as corresponding to different contractual structures.

In the past two decades, financial markets witnessed the implementation of three separate standards. The evolution of these standards was driven by a desire to achieve a consistent valuation, an objective still not fully achieved in Basel III regulatory documents and ISDA Master Agreements. The current standard is also plagued by various paradoxes related to DVA accounting which encourages anti-economical behaviour. Moreover, these paradoxes place an insurmountable limitation to the ability of replicating dynamically the price process of derivative transactions, an impossibility that hampers the role of banks as financial intermediaries.

The first proposal for a consistent valuation framework for CVA and DVA involves including a first-to-default clause in default protection contracts. This modification of the standard however represents a substantial deviation from the unilateral CVA specification in Basel III. To remedy in part, we introduce a new notion of portable CVA which is consistent and also very close to unilateral CVA.

We next consider three structures based on margin lending and full collateralization. We conclude that the quadri-partite structure whereby two margin lenders provide hypothecs to ensure full collateralization at all times to cover the open mark-to-market of derivative transactions is promising. These structures do not require a bank to carry out CVA trading as market risk and counterparty spread risk are transferred to the counterparty. Pure default risk is retained by the margin lenders that can securitize it in a straightforward fashion without being hampered by CVA volatility.
 
Finally, we consider the problem of dynamic replication for derivative transactions between defaultable entities. We introduce a penta-partite structure which is a variation of the quadri-partite structure above with the addition of a Central Counterparty. In the penta-partite case, we conclude that, if a replication strategy between non-defaultable entities exists, then this strategy can be implemented also by a defaultable structurer until its default time and the cost of replication is the risk-free fair value, independently of the credit quality of the structurer.  
 
For more detailed references on consistent global valuation, on DVA accounting controversies, on Basel III, on the ISDA Master agreements and closeout protocols, on Collateralization, Wrong Way Risk, Gap risk, DVA hedging, the first-to-default clause, and examples across asset classes we refer to the CVA FAQ \cite{BrigoFAQ}.

\appendix

\section{Probabilistic Framework and Absence of Arbitrage}

We denote by $\tau_B$ and $\tau_C$ respectively the default times of the Bank  and Counterparty. We fix the portfolio time horizon $T \in \mathbb{R}^+$, in that all default-free cash flows will be assumed to be zero after $T$, and fix the risk neutral pricing model $(\Omega,\mathcal{G},\mathbb{Q})$, with a filtration $(\mathcal{G}_t)_{t \in \mathbb{R}^+}$ such that $\tau_B$ and $\tau_C$ are $\mathcal{G}$-stopping times. Given that all cash flows go to zero after $T$, we can limit ourselves to the finite stopping times $\min\{\tau_B,T\}$ and $\min\{\tau_C,T\}$ for all practical purposes.

We denote by ${\Bbb{E}}_t$ the conditional expectation under $\mathbb{Q}$ given $\mathcal{G}_t$, and by ${\Bbb{E}}_{\tau_i}$ the conditional expectation under $\mathbb{Q}$ given the stopped filtration $\mathcal{G}_{\tau_i}$. We exclude the possibility of simultaneous defaults, assuming that
$\mathbb{Q}(\tau_B = \tau_C) = 0$. 

The pivotal theoretical result for asset valuation is the Fundamental Theorem of Finance, formulated in 1931 by Bruno de Finetti \cite{BDF1931}, see also \cite{AL2009} for a modern account. The mathematical expression of the principle of arbitrage freedom can be stated as a set
of linear inequalities. The Fundamental Theorem indicates how to find all solutions to these equations and how to express them via transition probabilities which can be estimated, or implied, from market prices. This framework has been formulated under a discrete setup, and the notions of replication and hedging strategy play no role in the related characterization of arbitrage freedom. The evolution of this result has been considered for example by \cite{Dalang}, up to the most modern results on arbitrage free theory such as \cite{DelbaenSchachermayer}.  It is debatable whether the continuous framework is really needed or adds clarity for a realistic representation of no arbitrage conditions in actual markets, see also \cite{AL2009} and \cite{AGW2} for a related discussion. Nonetheless, for the mainstream purposes of this paper, we stick to the standard continuous time version of arbitrage freedom as formulated for example in  \cite{HarrisonKreps} and \cite{HarrisonPliska}, and specialized to the credit context in \cite{BieleckiRutkowski2002}, to which we refer for more details on arbitrage freedom under default risk. 

\bigskip

\bigskip

\bibliographystyle{amsplain}

\bibliography{restructuring}


\end{document}